\documentclass[twocolumn, prx, superscriptaddress]{revtex4-2}

\usepackage{lipsum}
\usepackage{amsfonts, amssymb, amsmath}
\usepackage{bbold}
\usepackage{graphicx}
\usepackage{amsmath, esint}
\usepackage{bm}
\usepackage{xcolor, calc}
\usepackage{blindtext}
\usepackage{natbib}
\usepackage{float}
\usepackage{comment}
\usepackage{hyperref}
\usepackage[capitalise]{cleveref}
\usepackage{physics}

\usepackage{contour}
\DeclareMathAlphabet{\mathpzc}{OT1}{pzc}{m}{it}

\def\#{\cdot}

\date{March 2022}

\begin{document}
	
	\setlength{\parindent}{0pt}  
	\title{Macroscopic Hyperpolarization Enhanced with Quantum Optimal Control}
	
	\author{Alastair Marshall}
	\thanks{These three authors contributed equally to this work}
	
	\affiliation{NVision Imaging Technologies GmbH, D-89081 Ulm, Germany}
	\affiliation{Institute for Quantum Optics (IQO) and Center for Integrated Quantum Science and Technology (IQST), Albert-Einstein-Allee 11, Universit{\"a}t Ulm, D-89081 Ulm, Germany}
	
	\author{Thomas Reisser}
	\thanks{These three authors contributed equally to this work}
	
	\affiliation{Forschungszentrum J{\"u}lich GmbH, Peter Gr{\"u}nberg Institute -  Quantum Control (PGI-8), D-52425 J{\"u}lich, Germany}
	\affiliation{Institute for Theoretical Physics, University of Cologne, D-50937 Cologne, Germany}
	
	\author{Phila Rembold}
	\thanks{These three authors contributed equally to this work}
	\affiliation{Forschungszentrum J{\"u}lich GmbH, Peter Gr{\"u}nberg Institute -  Quantum Control (PGI-8), D-52425 J{\"u}lich, Germany}
	\affiliation{Institute for Theoretical Physics, University of Cologne, D-50937 Cologne, Germany}
	\affiliation{Dipartimento di Fisica e Astronomia ``G. Galilei”, Universit{\`a} degli Studi di Padova, I-35131 Padua, Italy}
	\affiliation{Istituto Nazionale di Fisica Nucleare (INFN), Sezione di Padova, I-35131 Padua, Italy}

	\author{Christoph M{\"u}ller}
	\affiliation{NVision Imaging Technologies GmbH, D-89081 Ulm, Germany}
	
	\author{Jochen Scheuer}
	\affiliation{NVision Imaging Technologies GmbH, D-89081 Ulm, Germany}
	
	\author{Martin Gierse}
	\affiliation{NVision Imaging Technologies GmbH, D-89081 Ulm, Germany}
	\affiliation{Institute for Quantum Optics (IQO) and Center for Integrated Quantum Science and Technology (IQST), Albert-Einstein-Allee 11, Universit{\"a}t Ulm, D-89081 Ulm, Germany}
	
	\author{Tim Eichhorn}
	\affiliation{NVision Imaging Technologies GmbH, D-89081 Ulm, Germany}
	
	\author{Jakob M. Steiner}
	\affiliation{NVision Imaging Technologies GmbH, D-89081 Ulm, Germany}
	\affiliation{Paul Scherrer Institute, CH-5232 Villigen PSI, Switzerland}
	
	\author{Patrick Hautle}
	\affiliation{Paul Scherrer Institute, CH-5232 Villigen PSI, Switzerland}
	
	\author{Tommaso Calarco}
	\affiliation{Forschungszentrum J{\"u}lich GmbH, Peter Gr{\"u}nberg Institute -  Quantum Control (PGI-8), D-52425 J{\"u}lich, Germany}
	\affiliation{Institute for Theoretical Physics, University of Cologne, D-50937 Cologne, Germany}
	
	\author{Fedor Jelezko}
	\affiliation{Institute for Quantum Optics (IQO) and Center for Integrated Quantum Science and Technology (IQST), Albert-Einstein-Allee 11, Universit{\"a}t Ulm, D-89081 Ulm, Germany}
	
	\author{Martin B. Plenio}
	\affiliation{Institute of Theoretical Physics (ITP) and Center for Integrated Quantum Science and Technology (IQST), Albert-Einstein-Allee 11, Universit{\"a}t Ulm, D-89081 Ulm, Germany}
	
	\author{Simone Montangero}
	\affiliation{Dipartimento di Fisica e Astronomia ``G. Galilei”, Universit{\`a} degli Studi di Padova, I-35131 Padua, Italy}
	\affiliation{Istituto Nazionale di Fisica Nucleare (INFN), Sezione di Padova, I-35131 Padua, Italy}
	\affiliation{Padua Quantum Technologies Research Center, Universit{\`a} degli Studi di Padova, I-35131 Padua, Italy}
	
	\author{Ilai Schwartz}
	\affiliation{NVision Imaging Technologies GmbH, D-89081 Ulm, Germany}
	
	\author{Matthias M. M{\"u}ller}
	\affiliation{Forschungszentrum J{\"u}lich GmbH, Peter Gr{\"u}nberg Institute -  Quantum Control (PGI-8), D-52425 J{\"u}lich, Germany}
	
	\author{Philipp Neumann}
	\affiliation{NVision Imaging Technologies GmbH, D-89081 Ulm, Germany}
	
	\begin{abstract}
		Hyperpolarization of nuclear spins enhances nuclear magnetic resonance signals, which play a key role for imaging and spectroscopy in the natural and life sciences. This signal amplification unlocks previously inaccessible techniques, such as metabolic imaging of cancer cells. In this work, electron spins from the photoexcited triplet state of pentacene-doped naphthalene crystals are used to polarize surrounding protons. As existing strategies are rendered less effective by experimental constraints, they are replaced with optimal control pulses designed with RedCRAB. In contrast to previous optimal control approaches, which consider one or two effective nuclei, this closed-loop optimization is macroscopic. A 28\% improvement in signal and 15\% faster polarization rate is observed. Additionally, a strategy called \textbf{A}utonomously-optimized \textbf{R}epeated L\textbf{I}near \textbf{S}w\textbf{E}ep (ARISE) is introduced to efficiently tailor existing hyperpolarization sequences in the presence of experimental uncertainty to enhance their performance. ARISE is expected to be broadly applicable in many experimental settings.
	\end{abstract}

	\maketitle
	
	\section{Introduction}
	
	\subsection{Dynamic Nuclear Polarization}
	Sensitive Nuclear Magnetic Resonance (NMR) spectroscopy and Magnetic Resonance Imaging (MRI) are key drivers in research areas from life sciences through material science to quantum computing. 
	The feasibility and sensitivity of such experiments critically depends on the polarization of the utilized spins.
	Dynamic Nuclear Polarization (DNP) techniques have been shown to increase NMR signals by multiple orders of magnitude~\cite{Larsen2003}, enabling previously inaccessible imaging techniques~\cite{Wang2019}. 
	DNP transfers the polarization from highly polarized electron spins to a target species of nuclear spins~\cite{Tateishi2014} used for NMR protocols.
	Electron spins are polarized, for example, by thermalization at low temperatures and high magnetic fields or by optical polarization of atoms and suitable molecules in gases, liquids, and solids \cite{Lilburn2013,Albert1994,Optical_Polarization_of_Molecules,Optical_Pumping_of_Interacting_Spin_Systems,liu_photo-induced_2017,kingRoomtemperatureSituNuclear2015, London2013, Scheuer_2016, Larsen2003}.
	\subsection{Pentacene-Doped Naphthalene Crystals as a Target}
\begin{figure*}
		\centering
		\includegraphics[width = \textwidth]{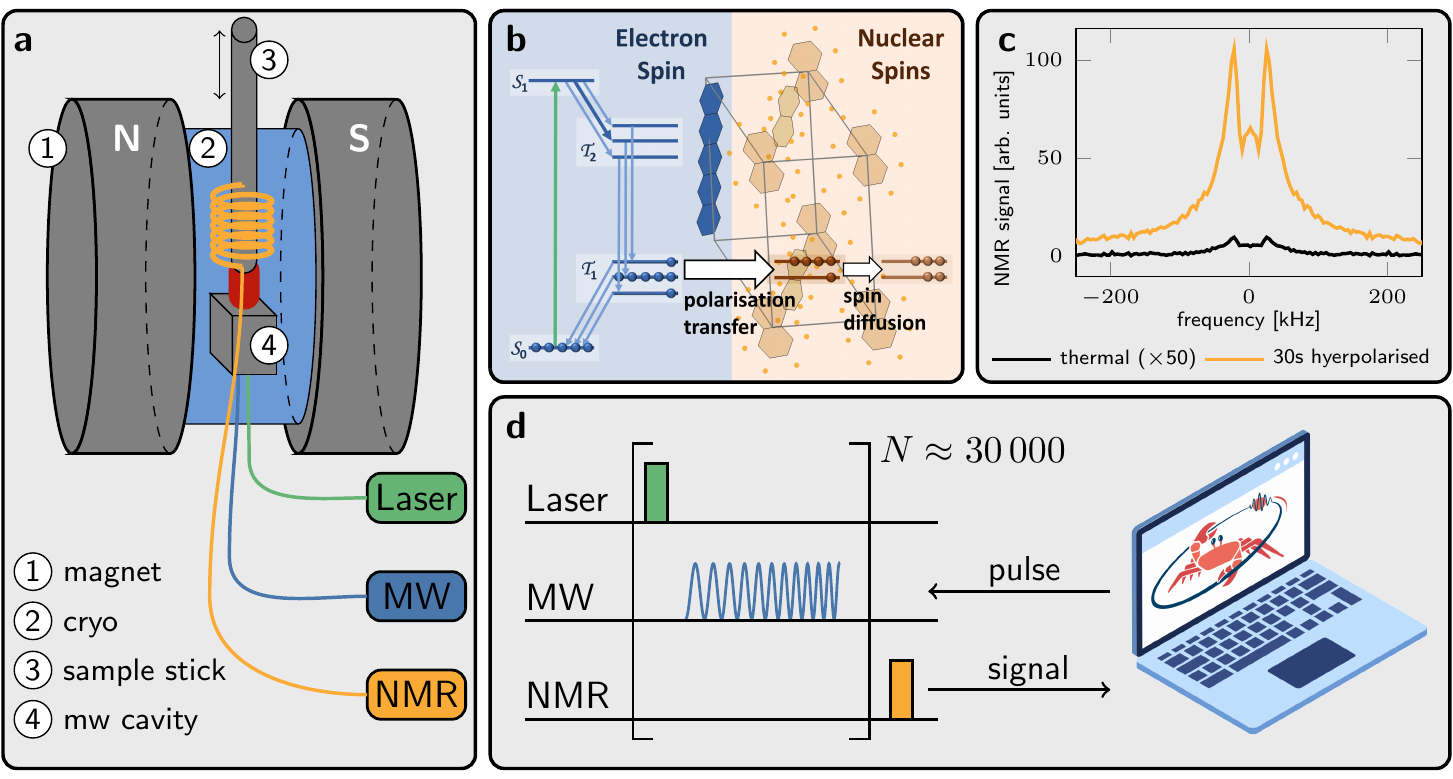}
		\caption{\textbf{Experimental realization.} a) Schematic of the in-house polarizer device. The sample (red) is mounted on a sample stick, which allows moving it between a MW cavity and an NMR coil inside a magnetic field at cryogenic temperatures. The laser is coupled to the system via an optical fiber. b) Level scheme of pentacene (electron spin) and naphthalene-based protons (nuclear spins), including the effect of laser excitation (green) and natural decay (blue). Spin diffusion to external nuclear spins~\cite{Pinon2017} is indicated. c) 30\,s of hyperpolarization show a clear signal enhancement compared to a 1h thermal build-up, the thermal signal is scaled by a factor of 50 to emphasize its faint polarization peaks. d) Schematic of the pulse sequence, consisting of a laser pulse for electron spin initialization and a MW pulse (of variable frequency) for polarization transfer to the nuclear spins. This basic block is repeated with a repetition rate of 1\,kHz (i.e. 30,000 repetitions are performed in 30\,s). After the polarization, the sample is shuttled into the NMR coil, where the magnetization is measured. An integral over the detected proton polarization is passed to RedCRAB, which provides the shape of the next MW pulse.}
		\label{fig:setup}
	\end{figure*}

	In this work, the electron spins of photoexcited triplet states in pentacene are used as the source of polarization, and the proton spins of naphthalene as the target. This system, shown in \cref{fig:setup}b, exhibits unique properties. In its ground state, the electron spin is in a singlet state and therefore the host crystal is free of paramagnetic defects. Consequently, proton relaxation times of 50 hours and above have been demonstrated at 77\,K and 0.5\,T~\cite{QUAN201922}. 
	
	In its metastable triplet state, the pentacene molecule exhibits a highly polarized electron spin with favorable lifetimes. Together with surrounding nuclear spins, this forms a central spin system that resembles other well-known systems like NV centers in diamond or phosphorous in silicon. This quantum resource for DNP leads to record values of 80\% proton polarization~\cite{quan2019novel}, which amounts to a polarized proton concentration of 50\,M. Exemplary applications of these nuclear spin polarized crystals are portable neutron spin filters in neutron scattering experiments~\cite{Quan_2019,QUAN201922} and polarization agents for NMR spectroscopy~\cite{eichhorn2021hyperpolarized,Kouno2019}.
	
	Under typical operating conditions (e.g., high magnetic field), electron and nuclear spins are mutually off resonant, prohibiting direct polarization transfer. Advanced spin control methods, like DNP, can be used to transfer polarization. Real-world experimental constraints like material quality, field inhomogeneities, and limited power and bandwidth commonly impair the ideal performance of existing DNP methods. Under such constraints, the maximum achievable polarization is reduced and the time to reach a certain polarization increases.
	
	\subsection{Existing DNP Strategies}
	In the case where the heterogeneity among spins is sufficiently small, they are all equally well controllable. As a result, techniques like nuclear orientation via electron-spin locking (NOVEL)~\cite{HENSTRA1988b, BrunnerFritschHausser1987} can be employed to transfer polarization.
	As the environmental complexity and inhomogeneity increases, other techniques are needed. 
	Transferring polarization while counteracting a broad Electron Spin Resonance (ESR) is done with the so-called “Integrated Solid Effect'' (ISE)~\cite{Henstra1988, Henstra1990}:
	After the electron spin initialization, either a linear magnetic field sweep is performed, while the sample is driven by a constant microwave (MW) field $B_1$, or a linear MW frequency sweep is performed at a static magnetic field.
	The ISE method is notable for both its simplicity and robustness and has been shown to reach up to 80\% total nuclear polarization in naphthalene under optimized conditions (e.g., liquid He cooling, sample quality)~\cite{quan2019novel}. It has also been applied to NV centers in diamond at room temperature~\cite{Miyanishi_2021,Schwartz_2018,Chen2015,Scheuer_2016}.
	While optimizing DNP sequences on a model, i.e., performing open-loop Quantum Optimal Control (QOC), is one method to recover some of their performance~\cite{Yuan2015,Pomplun2008,Pomplun2010}, another is to employ closed-loop QOC by allowing an algorithm to directly control the experiment (shown in \cref{fig:setup}d)~\cite{Brif2010,Glaser2015,Koch2016,Rembold2020,Mueller2021decade}. The latter approach is particularly appealing when the experimental setting is very complex or impossible to accurately model.
	Due to the complex molecular environment, coupled with experimental constraints, the system's true transfer function 
	is obscured, making accurate simulation difficult. This work represents the first time that closed-loop QOC has been applied to optimize DNP sequences.
	
	\subsection{Quantum Optimal Control}
	Many advances in quantum technology were only possible due to the design of sophisticated control strategies using methods of QOC~\cite{Brif2010,Glaser2015,Koch2016,Rembold2020,Mueller2021decade}. Established methods of QOC include gradient-based algorithms like GRAPE (gradient ascent pulse engineering) \cite{KhanejaGRAPE, DynamoPackage}, the Krotov algorithm~\cite{Konnov1999,KrotovPackage} or gradient-based algorithms based on automatic differentiation \cite{PhysRevA.95.042318}, as well as algorithms based on an expansion of the control pulse into a truncated basis like the dressed Chopped RAndom Basis (dCRAB) algorithm \cite{Doria2011,dressing_the_crab,Mueller2021decade}, typically coupled with direct search maximization algorithms.
	This pulse expansion ansatz can also be combined with the gradient approach~\cite{Machnes2018,Lucarelli2018,Soerensen2018}.
	The dCRAB algorithm is readily applicable to closed-loop control as it can be integrated directly with an experiment, allowing the user to treat the system as a black-box. For this purpose, the dCRAB algorithm was implemented in the QOC software packages Remote-dCRAB (RedCRAB) \cite{Zoller2018, Hoeb2017, Frank2017, Heck2018} and its open-source version Quantum Optimal Control Suite (QuOCS)~\cite{Rossignolo2021}. Recently, RedCRAB enabled automatic calibration of quantum gates~\cite{Frank2017} and robust sensing operations~\cite{Oshnik2021} with NV centers in diamond, optimization of BEC creation in ultracold atoms~\cite{Heck2018} and the creation of a 20-atom Schrödinger cat state with Rydberg atoms in an optical lattice~\cite{Omran2019}.

	In this work, the efficiency of the overall proton polarization process is increased by optimizing the DNP transfer process using closed-loop QOC. To guide the algorithm towards a solution which produces a strongly increased signal, it is helpful to provide a good initial guess. Consequently, a new multi-step protocol is introduced called \textbf{A}utonomously-optimized \textbf{R}epeated L\textbf{I}near \textbf{S}w\textbf{E}ep (ARISE).
	It provides a systematic approach to the improvement of ISE-like linear sweep DNP sequences in the presence of an unknown experimental transfer function.\\
	In \cref{sec:theory} the model used in the simulation is introduced. The measurement techniques are explained in more detail in \cref{sec:meas_technique} and a description of the experimental setup is given in \cref{sec:exp_realization}. Details of the simulation are given in \cref{sec:simulation}. The polarization results are introduced in \cref{sec:polresults} and a description of the new ARISE protocol is given \cref{sec:protocol}. In \cref{sec:discussion} the results of the paper are put into context.
	
	\section{Theory}
	\label{sec:theory}
	
	
	The spin system is modelled as an electron spin coupled to three nuclear spins. Both the electron and nuclear spins are considered to be spin-half particles. A strong, constant magnetic field $\boldsymbol{B}_0=B_0 \hat{z}$ is aligned along the long axis of the pentacene molecule, representing the $z$-axis. The total spin Hamiltonian of the electron can be written as,
	\begin{equation}\label{eq:Electron_Hamiltonian}
		\begin{split}
			H_{el}  = \frac{\hbar}{2} \omega_{0S} \sigma_z + \frac{\hbar^2}{4} \Bigg[ D \left( \sigma_z^2 - \frac{1}{3} \boldsymbol{\sigma}(\boldsymbol{\sigma}+1) \right) \\
			+ \; E (\sigma_x^2 - \sigma_y^2) \Bigg],
		\end{split}
	\end{equation}
	where $\boldsymbol{\sigma}=\{\sigma_x,\sigma_y,\sigma_z\}$ are the Pauli matrices. The Zeeman interaction is described by ${\omega_{0S} = - \gamma_S {B}_0}$, where $\gamma_S$ is the electron spin's gyromagnetic ratio. The factors $D$ and $E$ correspond to the zero-field splitting~\cite{Wenckebach_II_2014, Strien_1980}. The exact transition frequency is determined experimentally, and the magnetic field is aligned such that the splitting is symmetric.
	
	As shown in \cref{fig:setup}b the electron spin of the pentacene molecule is excited to the $\mathcal{S}_1$ state with a short laser pulse. From there it decays to the triplet states $\mathpzc{T}_2$ and subsequently $\mathpzc{T}_1$ via inter-system crossing (ISC)~\cite{Wenckebach_II_2014}. $\mathpzc{T}_1$ then couples to the nuclear spins in the vicinity of the molecule.
	The three states of the triplet correspond to spin quantum numbers $m_s = 0$ and $m_s = \pm 1$. An external magnetic field induces a Zeeman splitting of the $m_s = \pm 1$ levels, allowing for a two-level approximation.
	As the pentacene is deuterated, the resonances of the pentacene's own nuclear spins are shifted far enough from the other protons in the crystal that they can be neglected. The electron spin is assumed to have its origin at the center of the pentacene molecule. To extract the parallel and perpendicular dipolar coupling to the pentacene's electron spin, 574 protons of the nearest naphthalene molecules contained in the $3\times3\times3$ unit cells around the pentacene molecule are modelled.
	
	The driving field has a carrier frequency which is resonant with the electron spin  $\omega_{\text{res}} = D-\omega_{0S}$. Its amplitude $\Omega_{\text{ext}}(t)$ and phase $\varphi_{\text{ext}}(t)$ are modulated to control the system.  It is then transformed to the field inside the cavity $\Omega_{\text{int}}$ (the details are given below in \cref{eq:cavity}).\\
	Coupling between the electron and nuclear spins is described by the hyperfine interaction tensor $\textbf{A}^i$ with the nuclear spin indices $i = \{1, 2, 3\}$. A detuning $\Delta_{\text{es}}$ is introduced, describing the deviation of the field inside the cavity from the electron resonance frequency. In the rotating frame of the MW, and after applying the rotating wave approximation, the Hamiltonian is given by \cite{Wenckebach_I_2014}
	\begin{equation}\label{eq:Sys_Hamiltonian}
		\begin{split}
			H = \;  \hbar \Bigg( \Re[\Omega_{\text{int}}(t)] S_{x}
			+ \Im[\Omega_{\text{int}}(t)] S_{y}
			+ \Delta_{\text{es}} &S_{z} \\
			+ \, \omega_{\text{L}} \sum_{i=1}^{3} I^i_{z} 
			+ \sum_{i=1}^{3} \textbf{S} \# \textbf{A}^i \# \textbf{I}^i \Bigg),
		\end{split}
	\end{equation}
	where $\textbf{S} = \{S_x, S_y, S_z\}$ with $S_{k} = \frac{1}{2} \sigma_k \otimes \mathbb{1} \otimes \mathbb{1} \otimes \mathbb{1}, \; (k\in\{x, y, z\})$ are the electron's spin operators. $\textbf{I}^i$ and $I^{i}_k$ are the equivalent operators for the nuclear spin with index $i$ and $\Omega_{\text{int}}$ is the complex, time-dependent field inside the cavity
	$\omega_{\text{L}}\approx 9.2\,$MHz corresponds to the Larmor frequency of the nuclei. The voltage signal, which is fed into the AWG, is given by 
	\begin{equation}
		V_\text{ext} = V(t) \cos \big( (\omega_\text{res} + \Delta_\text{cs})\, t + \varphi_\text{ext}(t)\big).
	\end{equation}
	The conversion between $V(t)$ and $\Omega_{\text{ext}}(t)$ is determined directly from experimental data. The phase modulation $\varphi_{\text{ext}}(t)$ can be translated into the drive detuning $\Delta(t) = \dot{\varphi}_{\text{ext}}(t)$.
	
	The effect of the cavity on the external driving field is characterized by the cavity response factor $\gamma_{\text{cav}}$ and given by the differential equation
	\begin{equation}\label{eq:cavity}
		\begin{split}
			\frac{\partial}{\partial t} \Omega_{\text{int}}(t) = \gamma_{\text{cav}} \left( \Omega_{\text{ext}}(t) \cdot e^{-i\varphi_{\text{ext}}(t)} - \Omega_{\text{int}}(t) \right) \\
			- \, i \Delta_{\text{cs}} \Omega_{\text{int}},
		\end{split}
	\end{equation}
	where $\Delta_{\text{cs}}$ describes the constant detuning of the cavity from the resonance of the electron spin transition frequency~\cite{Cavity_Model}.
	
	In the secular approximation~\cite{Can_2015}, only the dominant coupling terms along $z$ are kept, giving
	\begin{equation}\label{eq:Hyperfine_Coupling}
		\textbf{S} \# \textbf{A}^i \# \textbf{I}^i \approx S_z \# \left( A_{zx}^i I_x^i + A_{zy}^i I_y^i + A_{zz}^i I_z^i \right).
	\end{equation}
	They are calculated for the respective position of the nucleus in the crystal structure by considering a purely dipolar interaction~\cite{Henstra_Thesis}.
	
	The static detuning values $\Delta_{\text{es}}$ for the Hamiltonian shown in \cref{eq:Sys_Hamiltonian} are drawn from a normal distribution with a full width half maximum (FWHM) of 10\,MHz to mimic additional frequency shifts due to magnetic field inhomogeneities and other impurities.
	
	The system description includes the dephasing of the electron spin via a Lindblad operator $R_{1} = \sqrt{\frac{\Gamma_{\text{el}}}{2}}\, S_{z}$, where $\Gamma_{\text{el}}$ is the dephasing rate~\cite{Lindblad_Marquardt}.
	The only electron states that interact with surrounding nuclei are the $\ket{0}$ and $\ket{1}$ states in the $\mathpzc{T}_1$ triplet (see \cref{fig:setup}). To account for the decay from $\mathpzc{T}_1$ to $\mathcal{S}_0$, the model includes a shelf state, which does not interact with the drive. It is only coupled via the loss rates $\Gamma_{\text{loss}, 0}$ and $\Gamma_{\text{loss}, 1}$ from the respective triplet states. The corresponding Lindbladians are given by $R_{2} = \sqrt{\Gamma_{\text{loss}, 0}} \,\sigma_{-, 0}$ and $R_{3} = \sqrt{\Gamma_{\text{loss}, 1}} \,\sigma_{-, 1}$ with $\sigma_{-, 0} = \ket{s} \bra{0}$ and $\sigma_{-, 1} = \ket{s} \bra{1}$.
	
	The evolution of the density matrix is then solved using the Lindblad master equation
	
	\begin{equation}\label{eq:LME}
		\dot{\rho} = - \frac{i}{\hbar} \left[ H, \rho \right] + \sum_{j = 1, 2, 3} \left( R_j \rho R_j^{\dagger} - \frac{1}{2} R_j^{\dagger} R_j \rho - \frac{1}{2} \rho R_j^{\dagger} R_j \right) \; ,
	\end{equation}
	where $\rho$ is the density matrix of the system.

	\section{Methods} 
	\label{sec:methods}
	
	A naphthalene crystal doped with pentacene-d14 grown in-house provides the electron spin system used as the polarization target. The crystal is placed in a magnetic field of 230\,mT at around 130\,K. This temperature is chosen because this is the limit of the polarizer. A series of $\sim 500\,$mW, $600\,$ns laser pulses initialize the pentacene molecules into their metastable spin-polarized triplet state $\mathpzc{T}_2$ (see \cref{fig:setup}a) via the singlet state, $\mathcal{S}_1$ from which it decays to the lower $\mathpzc{T}_1$ triplet via ISC \cite{Wenckebach_II_2014}. The pulse repetition rate is fixed at 1\,kHz. Depending on the occupation of the $\mathpzc{T}_1$-states, the pentacene returns to its ground state $\mathcal{S}_0$ after 80-180\,$\mu$s. 
	During this intermediate time, a MW DNP sequence transfers electron spin polarization to the densely packed proximal proton spins. 
	Strong dipolar coupling among protons distributes polarization throughout the entire crystal via spin diffusion. The macroscopic proton polarization is measured via NMR spectroscopy after a 30\,s buildup (see \cref{fig:setup}c).
	
	\begin{figure} 
		\centering
		\includegraphics[width = 0.47\textwidth]{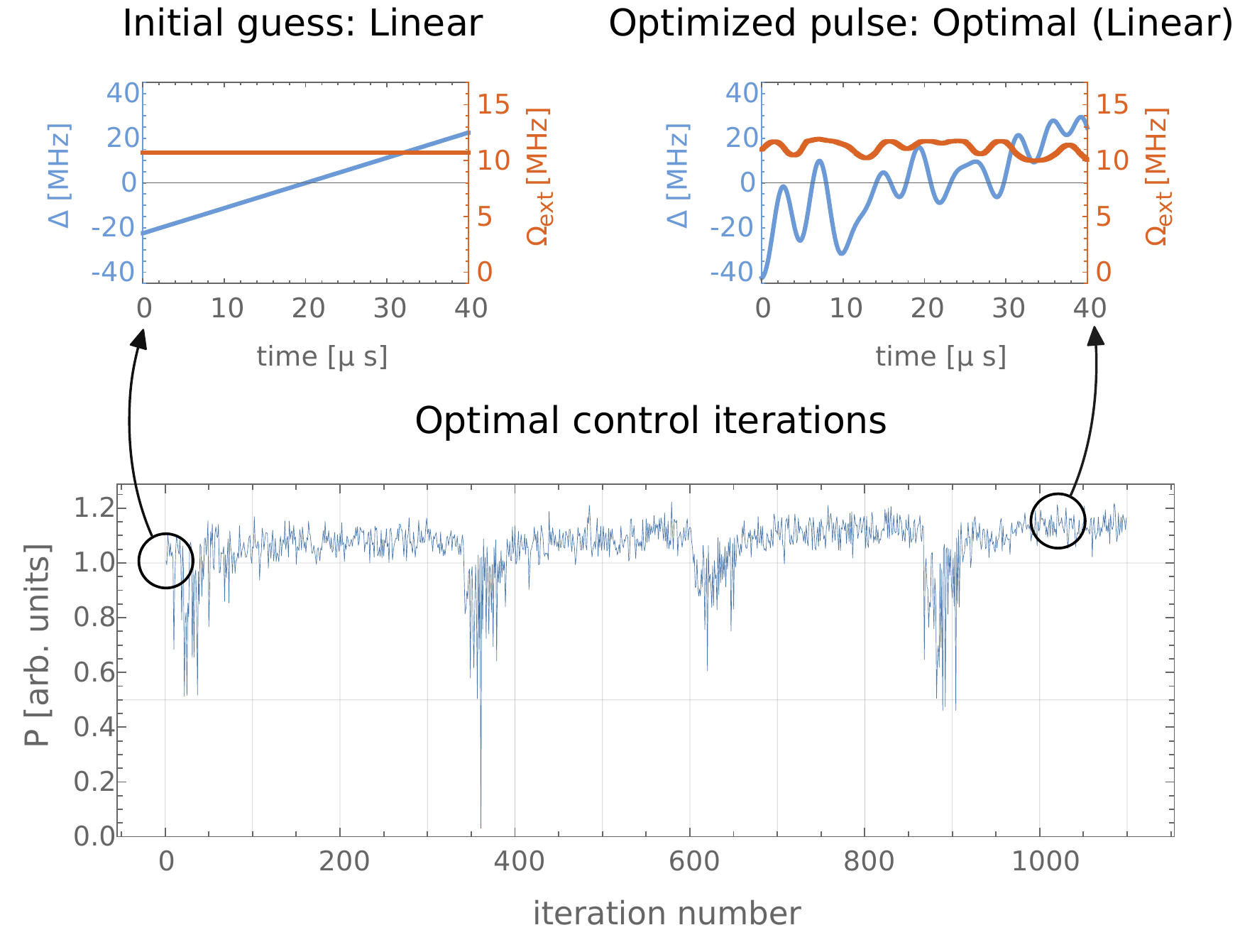}
		\caption{\textbf{Optimization procedure.} Comparison between the linear sweep pulse (\textit{top left}) and the pulse after optimization (\textit{top right}). The MW amplitudes are shown in orange and the detuning from cavity resonance in blue, representing the control pulses, which are shaped during the optimization. (\textit{bottom})~ The change of the FoM over different sub- and super-iterations during the optimization.}
		\label{fig:convergence}
	\end{figure}
	
	Starting from an external linear sweep (similar to ISE), and altering the amplitude and phase of the MW pulse using RedCRAB, this polarization is optimized (for more details see \cref{sec:protocol}). The optimization starting from a linear sweep is shown in \cref{fig:convergence}. Closed-loop QOC implicitly accounts for all experimental conditions influencing the transfer from optically initialized pentacene spins to macroscopic polarization. These could be, for example, the different lifetimes in the metastable state for different electron spin states or variations of the pentacence lifetimes throughout the crystal which might impact the final polarization.
	
	Another potential source of uncertainty could be the strong variation of couplings between the electron and surrounding proton spins or the distribution of polarization via spin diffusion during and after the MW DNP sequence. 
	Fluctuations in the experimental setup might also play a role, as bandwidth limitations of microwave equipment, spatial and spectral MW field variation inside the MW resonator, and spatial variation of laser intensity can also impact the efficiency.
	Additionally, it is challenging to control the amplitude of the MW field while its frequency is scanned across the resonance.
	Black-box (closed-loop) QOC does not directly incorporate variations in these parameters, whose role in polarization transfer is not understood, but if they play a role it can be captured by such an optimization.
	Importantly, some of these influences are very hard to predict theoretically.
	
	\subsection{Measurement Technique}
	\label{sec:meas_technique}
	The crystal grown in-house is cleaved along the $ab$ crystallographic plane, and mounted on a sample holder (see \cref{fig:setup}a) oriented along the $b$ crystallographic axis. The sample holder is then attached to a motorized stage, enabling it to be shuttled into the MW cavity, where it is cooled to 130\,K.
	Optically Detected Magnetic Resonance (ODMR) can be observed in pentacene-doped naphthalene crystals, where the triplet state is created using a 556$\,$nm laser pulse. By observing the fluorescence of the crystal under constant MW illumination, while changing the magnetic field, the electron spin resonance of pentacene can be found.
	The high-field transition of the pentacene triplet is used for both alignment and polarization. The crystal is then aligned by monitoring the ODMR spectrum while the sample is rotated, the best alignment is found when the resonance field is maximized. 
	Rabi oscillations are observed with a maximum Rabi frequency of 19.3$\,$MHz, by varying the duration of a resonant MW pulse.
	
	
	\subsection{Experimental Realization}
	\label{sec:exp_realization}
	
	The experimental results are obtained in an in-house polarizer device, shown in \cref{fig:setup}a, consisting of an optically accessible MW cavity inside an electromagnet operating at fields up to 800\,mT.
	The experimental sequence used to hyperpolarize the sample is shown in \cref{fig:setup}d. Within the MW cavity, photo-excited triplet states are created using a 556\,nm pulsed laser with a repetition rate of 1\,kHz delivering 0.35\,mJ optical power in a 600\,ns laser pulse every 1\,ms and setting the timing of experiments. The spin state is then manipulated using a MW pulse in between laser pulses. 
	Sophisticated pulse shapes can be sampled and uploaded to an Arbitrary Waveform Generator (AWG) using the experimental control software Qudi~\citep{Binder2017}. 
	The sample is attached to a holder that allows it to be shuttled into an NMR coil, which is located next to the MW cavity. Here, an NMR spectrometer (Magritek Kea$^2$) is used to read out the polarization of the proton spins using a 1Pulse measurement. This round-trip takes approximately 40 seconds from the pulse engineered by the RedCRAB software to the NMR measurement, and the optimizations typically ran for 12 hours.
	Integrating over the peak of the resulting NMR spectrum provides a relative estimate of the proton polarization. The RedCRAB algorithm is fed with this integrated signal and its estimated uncertainty to produce the next guess pulse.

	An optional 532\,nm continuous wave laser additionally allows the readout of the pentacene's electronic spin state optically; it is not used during closed-loop optimizations, but during the pentacene spin characterization experiments.
	Cooling of the sample is provided by a nitrogen gas flow system, which allows precise control of the temperature from 130\,K to above room temperature, as previously mentioned the experiments are carried out at this lower limit.

	\subsection{Simulation}
	\label{sec:simulation}
	The solutions to the differential equations in \cref{eq:LME} and \cref{eq:cavity} are calculated numerically using the DifferentialEquations.jl~\cite{DifferentialEquations.jl-2017} and other Julia packages~\cite{besard2018juliagpu, besard2019prototyping, Bromberger17, BenchmarkTools.jl-2016, VertexSafeGraphs.jl, Optim.jl-2018, rackauckas2017adaptive, LLVM.jl-2017, ma2021modelingtoolkit, AbstractAlgebra.jl-2017, quadgk, Zygote.jl-2018, Convex.jl-2014, MathOptInterface-2021, FFTW.jl-2005}. To obtain a realistic polarization build-up, \cref{eq:LME} is solved for and averaged over 1000 instances. For each instance, three random but distinct nuclei are picked from the 30 most strongly coupled nuclei and the detuning $\Delta_\text{es}$ is sampled from a Gaussian distribution. This way, mechanisms which are neglected in the common weighted sum single nucleus approximation are captured. Examples include the re-polarization of the electron spin through partially polarized nuclei or the redistribution of polarization from one nucleus to another. The mean over many runs with different coupling combinations takes into account the variety of couplings in the system with reasonable computational resources.

	The cavity response $\gamma_{\text{cav}}$ is determined by repeatedly applying constant external drive fields with different cavity detunings $\Delta_\text{cs}$, obtaining a photon count that corresponds to the electron state.
	
	\begin{figure}
		\centering
		\includegraphics[width = 0.45\textwidth]{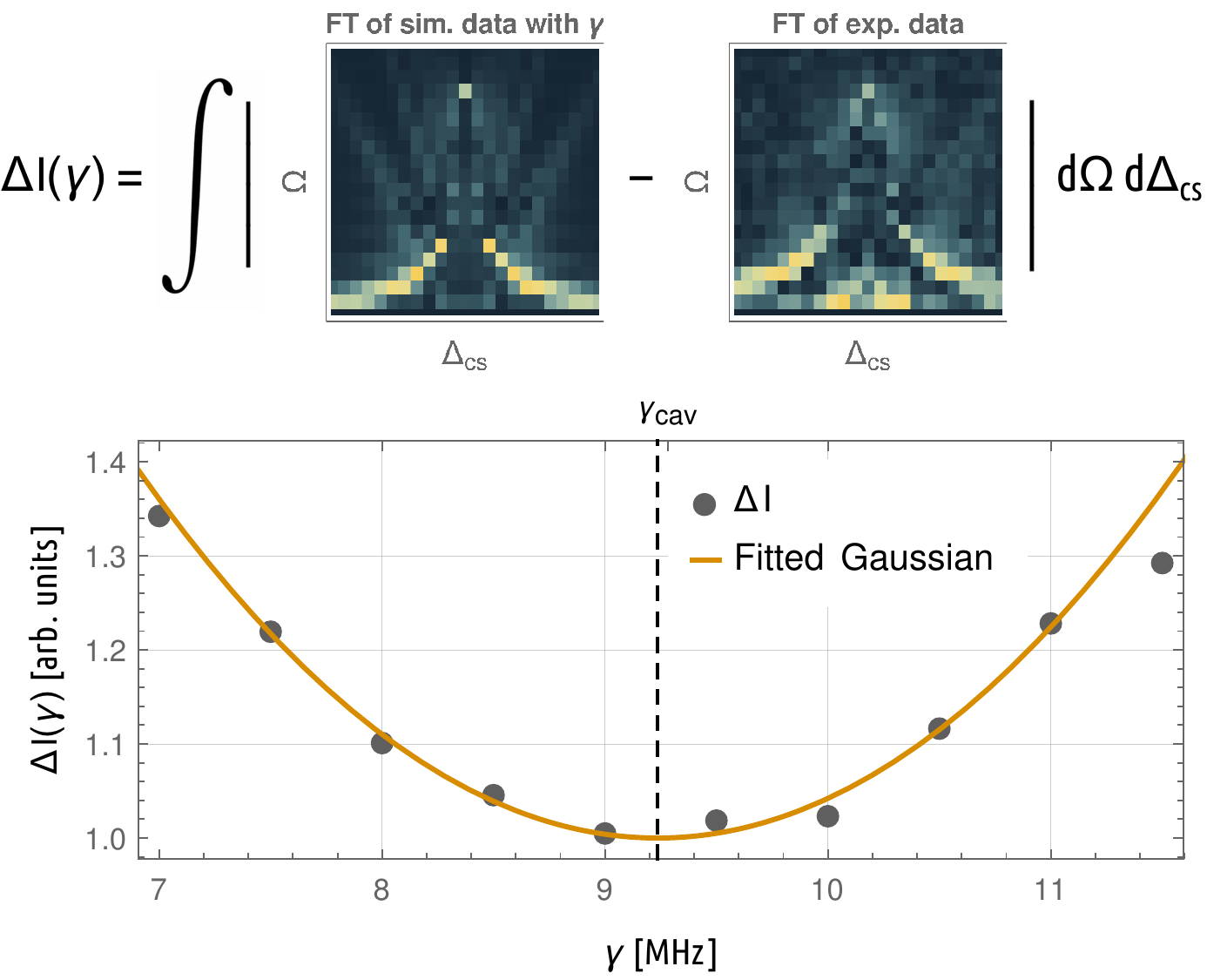}
		\caption{\textbf{Characterization of the set-up via response factor $\gamma$.} The agreement between experiment and simulation, $\Delta I$, is calculated via the Fourier transform $\Omega$ of the driven Rabi signal for varying cavity detunings (\textit{top}). The integral indicates the overlap of measured and simulated distribution, which is shown for different values of cavity response fed to the simulation in the bottom plot. The black line indicates the minimum of the error in overlap between simulation end experiment at $\gamma_{\text{cav}} = 9.24$\,MHz (\textit{bottom}).}
		\label{fig:gamma_heatmap_fit}
	\end{figure}

	$B_0$ is adjusted such that the spin always stays resonant with the drive frequency. Oscillations are recorded for times up to $0.6 \,\mu$s from the start of the drive pulse. The detuning is swept through a range of $\pm 25$\,MHz around the resonance. The maximum of the Fourier transform of the photon count then corresponds to the Rabi frequency $\Omega$ for a detuning $\Delta_\text{cs}$. The cavity dynamics are complex, leading to a response similar to the example shown at the top right of \cref{fig:gamma_heatmap_fit}.
	These measurements are modeled for an electron spin inside a cavity with a response factor $\gamma_{\text{cav}}$ between 5 and 14\,MHz (range suggested by response time based on the measured Q factor of the resonator using a spectrum analyzer). The resulting Fourier transforms are compared to the experimental values.
	The comparison was done by calculating the overlap of the normalized measurement and simulation grids, as shown in \cref{fig:gamma_heatmap_fit}. The minimum of the sum of the absolute difference between each grid point of  the measurement and simulation data is obtained by a Gaussian fit resulting in $\gamma_{\text{cav}} = 9.24$\,MHz.\\
	The values for the decoherence rate, $\Gamma_{\text{el}}$,  of the electron spin and the loss rates to the shelf state, $\Gamma_{\text{loss}, 0}$ and $\Gamma_{\text{loss}, -1}$, are found by performing a Hahn echo measurement and state-dependent lifetime measurements. For the dephasing time of the electron $1 / \Gamma_{\text{el}} = 10 \, \mu$s is obtained and the triplet-state decay times are measured to be $1 / \Gamma_{\text{loss}, 0} = 80 \, \mu$s and $1 / \Gamma_{\text{loss}, -1} = 180 \, \mu$s.\\

	
	\section{Results} 
	\subsection{Polarization Build-Up}
	\label{sec:polresults}
	
\begin{figure*}
		\makebox[\textwidth][c]{
			\includegraphics[width = \textwidth]{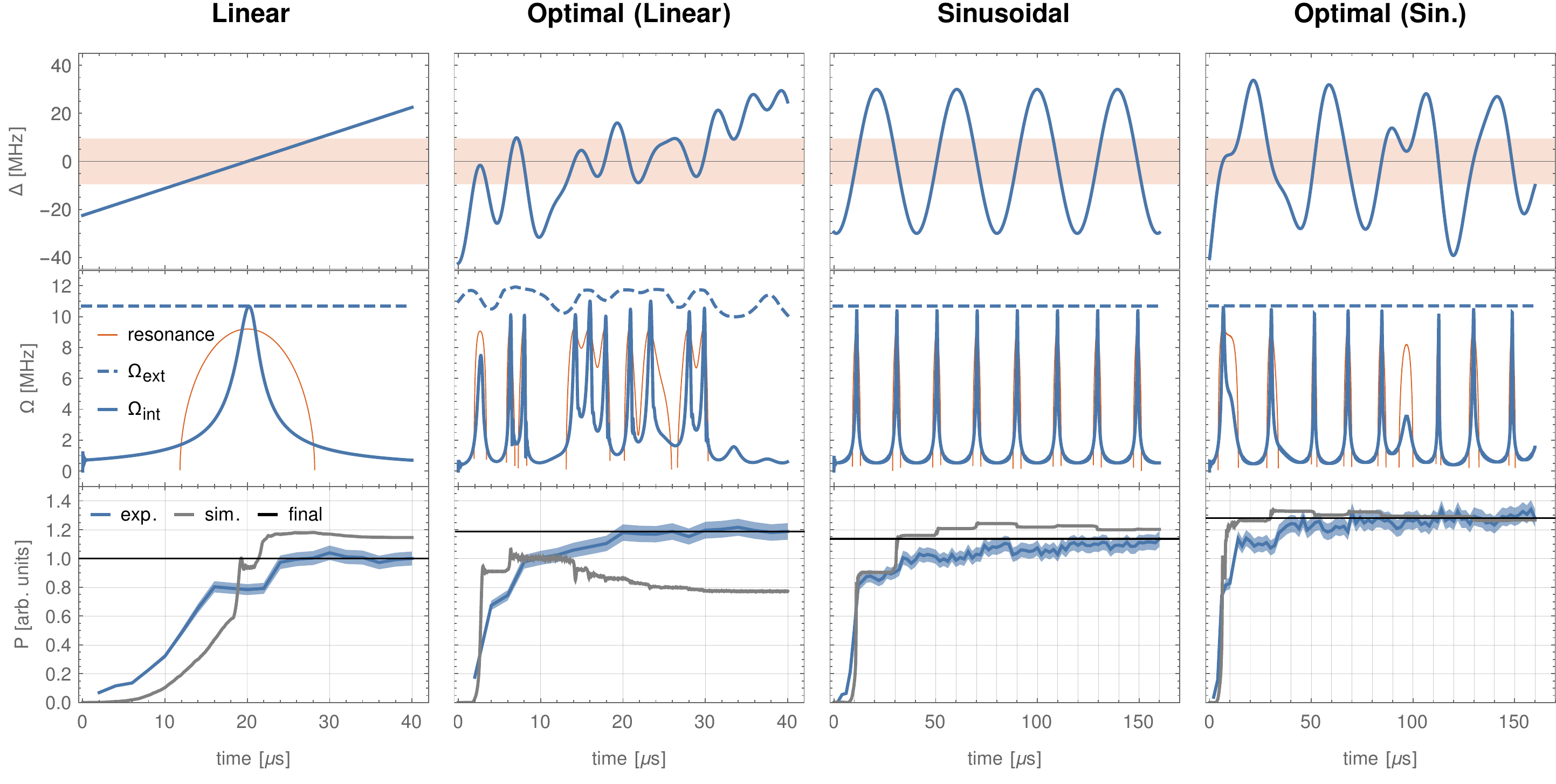}}
		\caption{\textbf{Implementation of a selection of polarization pulses.} From left to right, the figure presents the following MW pulse schemes on the resonator: The externally applied linear sweep, a linear sweep-based QOC-generated pulse, sinusoidal sweep of the detuning, a corresponding QOC-generated pulse. 
			The first row gives the detuning applied by the drive with respect to time. The orange area comprises the window in which the Hartmann-Hahn resonance condition lies ($\Delta^2=\omega_{0I}^2-\Omega^2$). The second row shows the Rabi frequency as applied externally (dashed), the field inside the cavity (solid blue), and the resonance condition for the given detuning (thin, orange). The last row shows how the polarization builds up over the course of the pulse. Experimental values (exp.) are shown in blue, theoretical values (sim. for simulation) in gray. The solid black line indicates the final polarization.}
		\label{fig:optimization_table}
	\end{figure*}

	The optimization directly adjusts the pulse phase for experimental convenience, but as the detuning modulation $\Delta$ contains the same information and connects directly to the system dynamics, it is displayed in \cref{fig:optimization_table} instead. The first row of \cref{fig:optimization_table} shows the detuning modulation $\Delta$ as a function of time during the pulse. 
	In the second row, the y-axis shows the amplitude $\Omega$ during the pulse for both the externally applied field and the internal cavity field.
	The Hartmann-Hahn resonance~\cite{Hartmann1962}, shown in orange, is calculated using the time-dependent detuning assuming the target spin is a proton.
	In the third row of \cref{fig:optimization_table}, experimental data and simulation are compared, showing how the polarization builds up during the pulse.
	
	The first initial guess pulse is an ISE-like linear sweep whose parameters (amplitude, sweep rate and duration) had already been manually tuned on the experimental setup. The externally applied amplitude is kept constant, while the frequency is swept across the resonance. This pulse serves as the benchmark against which the optimized pulses are compared both in the experiment and in the simulation.
	
	During the experiment, the polarization plateaus at the center of the pulse before continuing to rise in the second half of the pulse. The simulation results contain the same features. However, the plateau is shorter and the initial rise is slightly delayed compared to the experimental data. These differences might be caused by the lack of a full analytical model that extends the existing description of the cavity with distortions due to electronics and other components of the setup. It should be noted that polarization plateaus occur in the simulated results across all pulses when the detuning is above the resonance condition.

	The optimization of this guess results in the pulse labelled ``Optimal (Linear)'' which shows a relative polarization improvement of approximately 19\%. The evolution of the FoM during the search for an optimized pulse and the comparison between the initial and the optimized version are shown in \cref{fig:convergence}.
	
	The ``Optimal (Linear)'' pulse transfers the majority of its polarization during the first $10 \, \mu$s.
	It is notable that the detuning is oscillating during that time, and it crosses the cavity resonance several times. The algorithm slows the sweep down as the detuning approaches the resonance. Both features appear in multiple optimization outcomes. 
	Arguably, if some electron population is left untransferred, subsequent sweeps through the resonance in both directions (from positive to negative detuning and back) can serve as additional opportunities for the polarization of more weakly coupled nuclear spins.
	
	The relative speed-up of the polarization transfer during this pulse is visible in both the simulation, and the experiment. However, a divergence between the measurement and the simulation arises in the latter part of the pulse. This could be explained by the very fast oscillations in cavity field amplitude, which are not captured in full detail in the simulation, due to the unknown transfer function of the setup's electronics.
	
	After analyzing the effect of the amplitude and phase of the pulse independently (see \cref{app:a} for more details), the MW amplitude is kept constant in later optimizations.
	To further explore the idea that repeated sweeps through the resonance are beneficial, pulses with phase oscillations that use a range of frequencies and pulse durations are tested, see \cref{app:b} for more details.
	The ``Sinusoidal'' pulse shown in column three of \cref{fig:optimization_table} is guessed in this manner. It outperforms the linear sweep by approximately 14\%.
	%
	%
	The “Sinusoidal'' protocol in \cref{fig:optimization_table} then becomes the new guess pulse for the optimal control algorithm. After adding more frequency components, RedCRAB obtains the ``Optimal (Sin.)'' pulse. It outperforms all other pulses in both final polarization (approximately 28\% higher with respect to the linear sweep for the short build-up measurement) and polarization rate on our setup.
	When the pulse is significantly detuned from the resonance, as in the first 5\,$\mu$s, the energy gap between the spins is large and so very little polarization can be transferred. As this gap closes, the nuclei are more likely to be polarized and the pulse slows down to allow for an extended transfer period.
	The detuning “slow down'' was recreated with an analytical polynomial function. (For details see \cref{app:c}) The resulting ``Fitted Optimal'' pulse largely retains the polarization capability of the optimized pulse. The comparable efficiency corroborates that the “slow down''-feature contributes to the substantial polarization build-up during the first 30\,$\mu$s. This behavior is reminiscent of optimal adiabatic passages with Landau-Zener protocols and  optimal controlled crossings of quantum phase transitions. Both have been investigated in different theoretical and experimental scenarios~\cite{Stefanatos_2020,Roland2002,Caneva2011_Speeding,vanFrank2016,Malossi2013} providing a basis for further exploration.

	The simulated polarization of both, the ``Sinusoidal'' and the ``Optimal (Sin.)'' pulse, match the experimental data closely. Again, the initial step-plateau-step shape of the ``Sinusoidal'' is reduced after optimization. Almost all the performance gained by optimizing the pulses arise from the behavior during the first 40-50\,$\mu$s of the pulse. This is on a similar timescale as the electron's decay to the singlet state.
	In the simulation, the polarization of the longer pulses slowly decreases after ca. 80\,$\mu$s. This was not seen in the experiment, most likely due to spin diffusion. Many weakly coupled nuclear spins could lead to a slow distribution of polarization away from the electron spin.
	Spin diffusion is expected on a timescale of ca. 100\,$\mu$s according to calculations of the dipolar interaction strength between the protons~\cite{Pinon2017}.
	
\begin{figure}[ht]
		\centering
		\includegraphics[width = 0.45\textwidth]{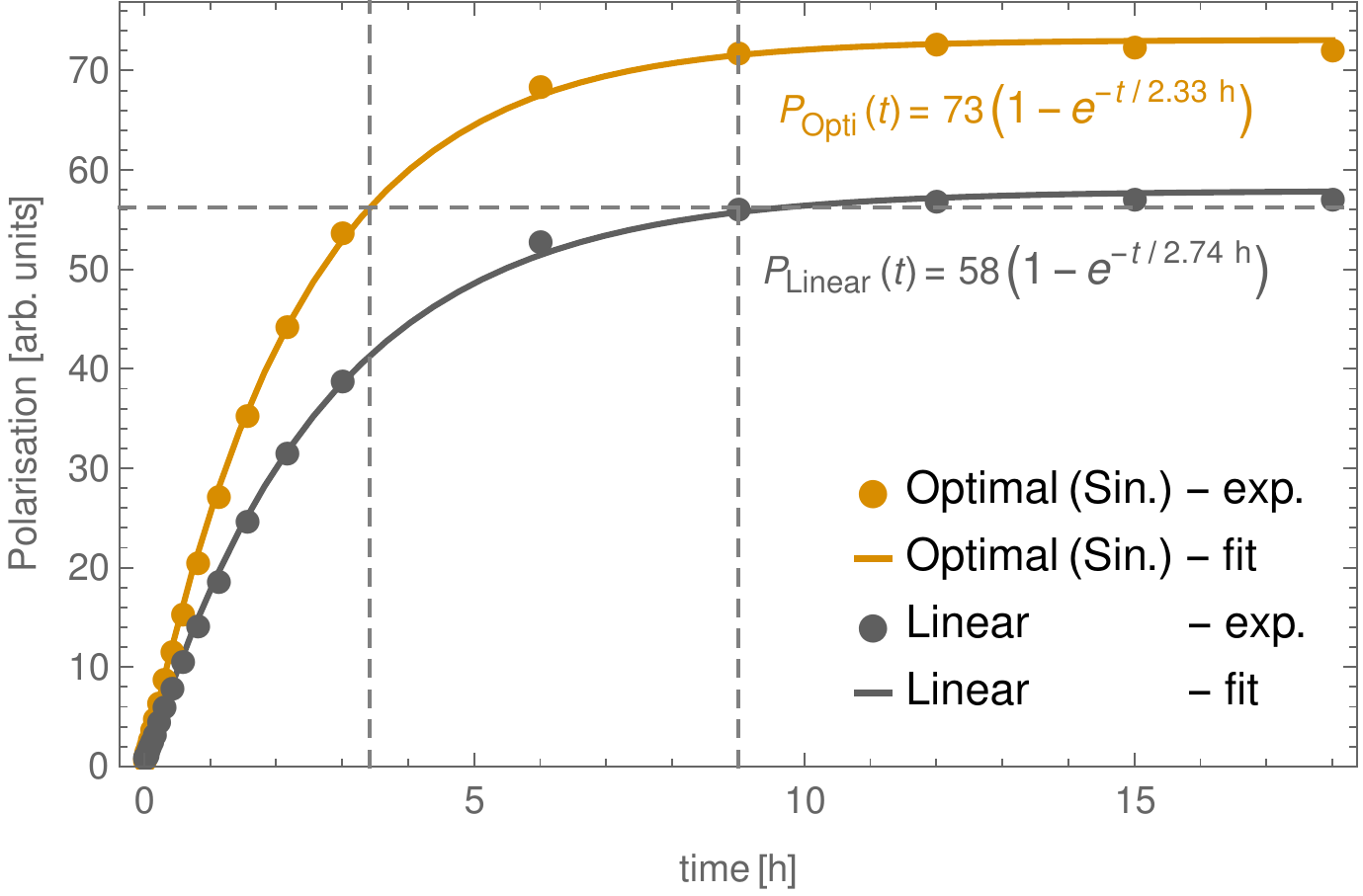}
		\caption{\textbf{Long-term polarization build-up.} Comparing the polarization build-up using the optimal settings for the linear sweep and the RedCRAB optimized pulse, shown in \cref{fig:optimization_table}. Using closed-loop optimal control, a higher final polarization  is reached in a shorter build-up time before saturation. The vertical dashed lines mark the times at which the polarization reaches 98\% of the maximum polarization of the linear sweep pulse. This level is reached in 3.35\,h using the optimized pulse, in contrast to 9\,h with the linear sweep. The formulas placed in the figure correspond to the exponential fits, $P_\text{Opti}$ and $P_\text{Linear}$, of the polarization build-up during the optimized pulse and the linear sweep, respectively.}
		\label{fig:build_up}
	\end{figure}

	The key result of the paper is shown in \cref{fig:build_up}, where the RedCRAB-optimized ``Optimal (Sin.)'' demonstrates two clear improvements over the linear sweep. Firstly, the magnitude of polarization increases by 26\% when using the optimized pulse.
	Secondly, the optimized pulse reaches, in only 3.35 hours, the same polarization that the linear sweep approach obtains after 9 hours of continuously repeating the protocol, making it a factor of 2.6 faster. These times correspond to $\sim$98\% of the maximum polarization of the linear sweep, which is within the error margin of the polarization measurements. 
	This allows for more than doubling the number of polarized crystals in a given time. Previously, crystals were left to polarize overnight to reach sufficiently high polarization. Using the optimal protocol, it is now possible, on the current experimental setup, to polarize several crystals per day. Due to its increased performance, the ``Optimal (Sin.)'' pulse was also applied as the hyperpolarization method of choice by Eichhorn et al.~\cite{eichhorn2021hyperpolarized}. In that paper, a bulk crystal polarization of 25\% is achieved using the optimized pulse.
	%
	
	The saturation of the polarization at this higher level is likely due to an equilibrium being reached between the polarizing sequence and the competing $T_1$ decay process of the nuclear spins~\cite{Pinon2017}.
	The lifetime of the nuclear states is measured to be approximately 3-4 hours under laser illumination, considering the specific values for the magnetic field $B_0$ and temperature. In this case, it is limited due to the laser illumination and MW fields causing, for example, additional heating.
	While this is the limit for the lifetime during polarization transfer, the polarization can be stored in the sample afterwards for much longer. Lifetimes between 50 hours \cite{eichhorn2021hyperpolarized} and 800 hours \cite{QUAN201922} have been reported for similar crystals at different temperatures and magnetic fields.
	
	Additionally, by examining the shape of the optimized pulse and fitting its main features, a simplified analytical function is obtained describing the pulse (shown in \cref{app:c}), labelled “Fitted Optimal''). This retains most of the enhanced performance of the optimized sweep.
	
	\subsection{ARISE}
	\label{sec:protocol}
	
	Generalizing the steps taken to achieve the results of the previous section, the \textbf{A}utonomously-optimized \textbf{R}epeated L\textbf{I}near \textbf{S}w\textbf{E}ep (ARISE) procedure is introduced. Each step provides a recipe for finding a good initial guess for the proceeding optimization. While this should be unimportant for an infinite-dimensionally parameterized optimization without limits and infinite measurement precision, in practice those restrictions apply, leading the algorithm to local instead of global optima. Despite the dCRAB algorithm's approach allowing it to escape local minima under certain circumstances, the optimization time is also drastically reduced if the initial guess is chosen carefully~\cite{Mueller2021decade}.
	The protocol consists of three steps:
	\begin{enumerate}
		\item \textbf{Tune the linear sweep.} Do a parameter search for the sweep range $\Delta_\text{max}$ and duration $t_\text{Linear}$ producing the most efficient polarization transfer.
		\item \textbf{Construct multi-sweep.} Set up a protocol which sweeps the detuning repeatedly between $\pm\Delta_\text{max}$ for $N_\text{osc}$ times with a period $\tau$. Do a parameter search for $N_\text{osc}$ and $\tau$, starting from $\tau=t_\text{Linear}$.
		\item \textbf{Apply quantum optimal control.} Search the full function space of the detuning $\Delta(t)$ using an optimal control algorithm. The initial guess is provided by the multi-sweep protocol from the previous step.
	\end{enumerate}
	In this work, steps one and two are accomplished through a simple parameter sweep. During the second step, the detuning is swept with the function $-\Delta_\text{max}\cos(2\pi t/\tau)$, however this could be replaced by linear sweeps. As the setup requires the pulse phase $\varphi_{\text{ext}}(t)$ as an input, all detunings are translated to phase modulations (see \cref{sec:theory}). In general, experimental feedback determines the best solution for the respective step. Here, it took the form of the proton NMR signal after 30,000 repetitions of the sequence. The third step is implemented using the RedCRAB software, which suggests different shapes for the phase of the pulse (see \cref{app:h}).
	
	\section{Discussion and Outlook} 
	\label{sec:discussion}
	
	The use of closed-loop optimal control provides a strategy for improving hyperpolarized NMR signals in complex experimental setups despite the unknown transfer function. A concern often raised about numerically optimized sequences is that they lose generality and only apply to a specific setup or sample.
	On the contrary, the ``Optimal (Sin.)'' has been successfully applied on different crystals with varying spin relaxation times across an extended period of time~\cite{eichhorn2021hyperpolarized}. It is now the gold-standard pulse in the lab.

	The combination of a 15\% faster polarization rate and a 26\% higher polarization level provides a factor of 2.6 reduction in the time taken to polarize crystals to within the margin of error of the previous method. As a result, multiple crystals can be polarized per day to be used in external hyperpolarization experiments~\cite{eichhorn2021hyperpolarized}.
	Furthermore, these improvements lead to higher levels of polarization in a shorter time, resulting in an overall polarization within the crystal of about 25\%~\cite{eichhorn2021hyperpolarized}. Such strongly polarized crystals are necessary to transfer polarization to external nuclear spins. By operating under liquid Helium conditions and with improved crystal quality, even higher values are anticipated~\cite{quan2019novel}.
	Mimicking the features of the optimized pulse by fitting an analytical function to it (as shown in \cref{app:c}) retains almost all the improved performance. It represents a good starting point not only for future optimization, but also for further investigations of the system dynamics. 

	A key feature of all the sequences that outperform the linear sweep is that they repeatedly sweep through the cavity resonance. Hence, the sequences are given an extra opportunity to transfer polarization, suggesting that the first sweep leaves some electron polarization untransferred.
	The simulation shows that the first sweep primarily transfers to one of the most strongly coupled nuclear spins, while subsequent sweeps redistribute the polarization to a wider range of couplings. This type of dynamics can only be accounted for when multiple nuclear spins are considered or the optimization is done on a macroscopic system.

	The ARISE protocol offers a starting point for future optimizations of DNP sequences using both open and closed-loop protocols. Inherently flexible, the protocol is easily customized to fit any number of setups, including the complex molecular environment seen here.
	
	In conclusion, the application of the ARISE protocol results in a 26\% improvement of the polarization level and 15\% faster polarization rate. Consequently, crystals were efficiently polarized to 25\% bulk proton polarization. These crystals were then used as the polarization source for an external hyperpolarization experiment~\cite{eichhorn2021hyperpolarized} which demonstrated strong transfer to external spins.

	\section{Acknowledgements}
	The authors would like to acknowledge receiving funding from the European Union’s Horizon 2020 research and innovation programme under the Marie Sk\l{}odowska-Curie \mbox{QuSCo} (N$^\circ$ 765267), EU Quantum Flagship projects \mbox{ASTERIQS} (N$^\circ$ 820394), \mbox{PASQuANS} (N$^\circ$ 817482) and the ERC Synergy grant \mbox{HyperQ} (N$^\circ$ 856432).
	
	
	
	
	\appendix
	\section{Amplitude vs. Phase Variation}
	\label{app:a}
	
	\begin{figure}[ht]
		\centering
		\includegraphics[width=0.48\textwidth]{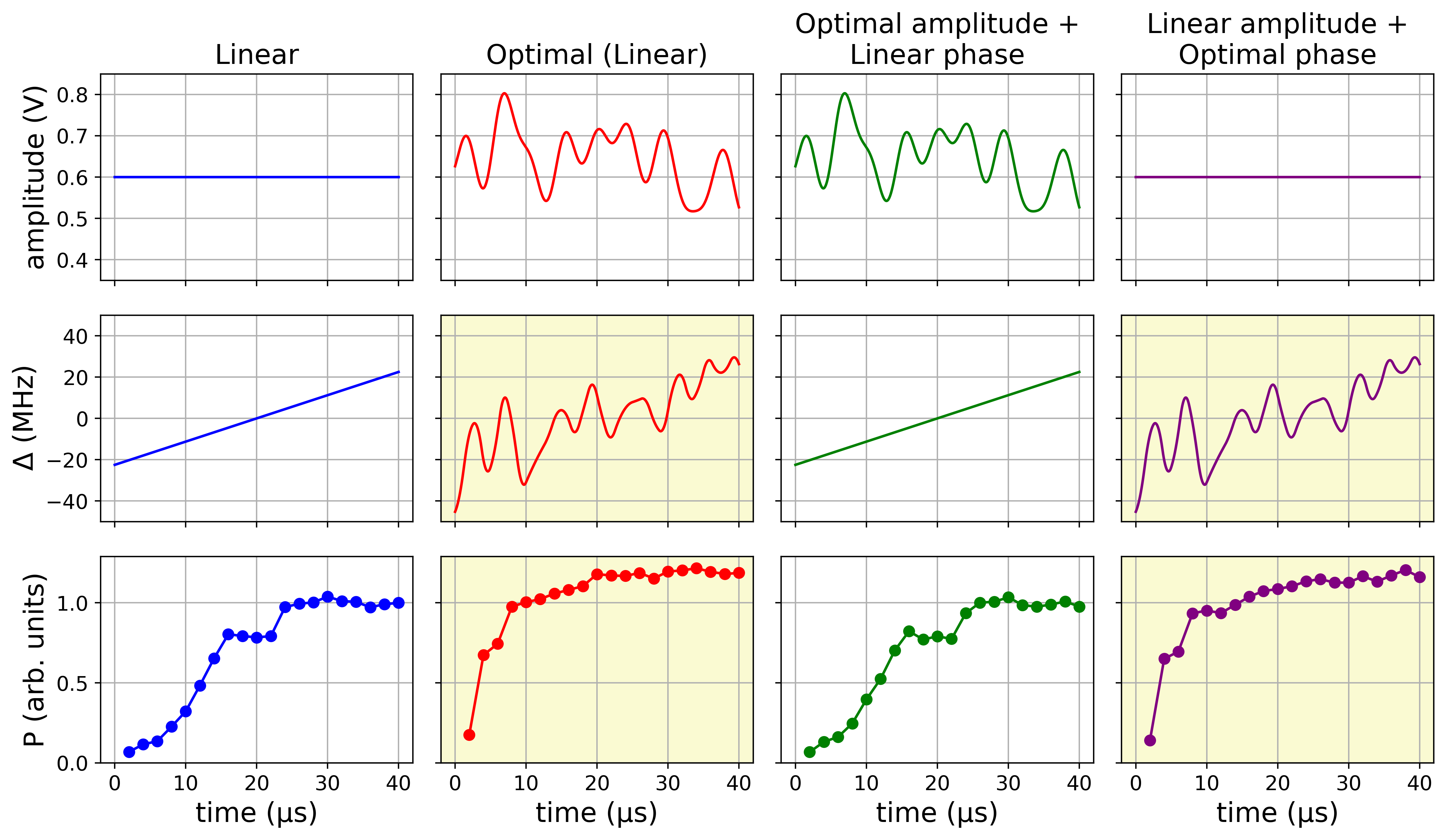}
		\caption{Amplitude vs phase variations: Comparing the linear sweep (Linear) to the corresponding QOC-generated pulse (Optimal (Linear)), as well as combinations of both. In the third and fourth column the amplitude array of one is combined with the phase array of the other (Optimal amplitude + Linear phase and Linear amplitude + Optimal phase). The increase in polarization over the optimization is caused solely by the changes in phase rather than amplitude (compare highlighted plots).}
		\label{fig:amp_vs_phase_var}
	\end{figure}
	
	To separate the respective effects of amplitude and phase modulation, these two parameters are investigated independently~(\cref{fig:amp_vs_phase_var}). The first test was a basic linear sweep and the Optimal (Linear) pulse, where it was observed that the optimized pulse leads to higher polarization. To test only the optimized amplitude modulation, the optimized phase is reset to the initial guess while the phase modulation was kept (Optimal amplitude + Linear phase). Similarly, to test the optimized phase modulation the amplitude is held constant, as in the linear sweep, and the optimized phase is applied (Linear amplitude + Optimal phase).
	
	By comparing those four pulses, it becomes clear that the amplitude modulation plays no role in the polarization transfer and only the pulses with optimally controlled phase modulation lead to enhanced polarization (highlighted in~\cref{fig:amp_vs_phase_var}) .
	Testing the phase of the optimized pulses with different constant MW amplitudes shows that using a higher MW amplitude always leads to better polarization transfer (see \cref{fig:comparison_increasing_amplitudes}).
	
	\section{Sine Oscillation Frequency Tests}
	\label{app:b}
	
	\begin{figure}[ht]
		\centering
		\includegraphics[width=0.40\textwidth]{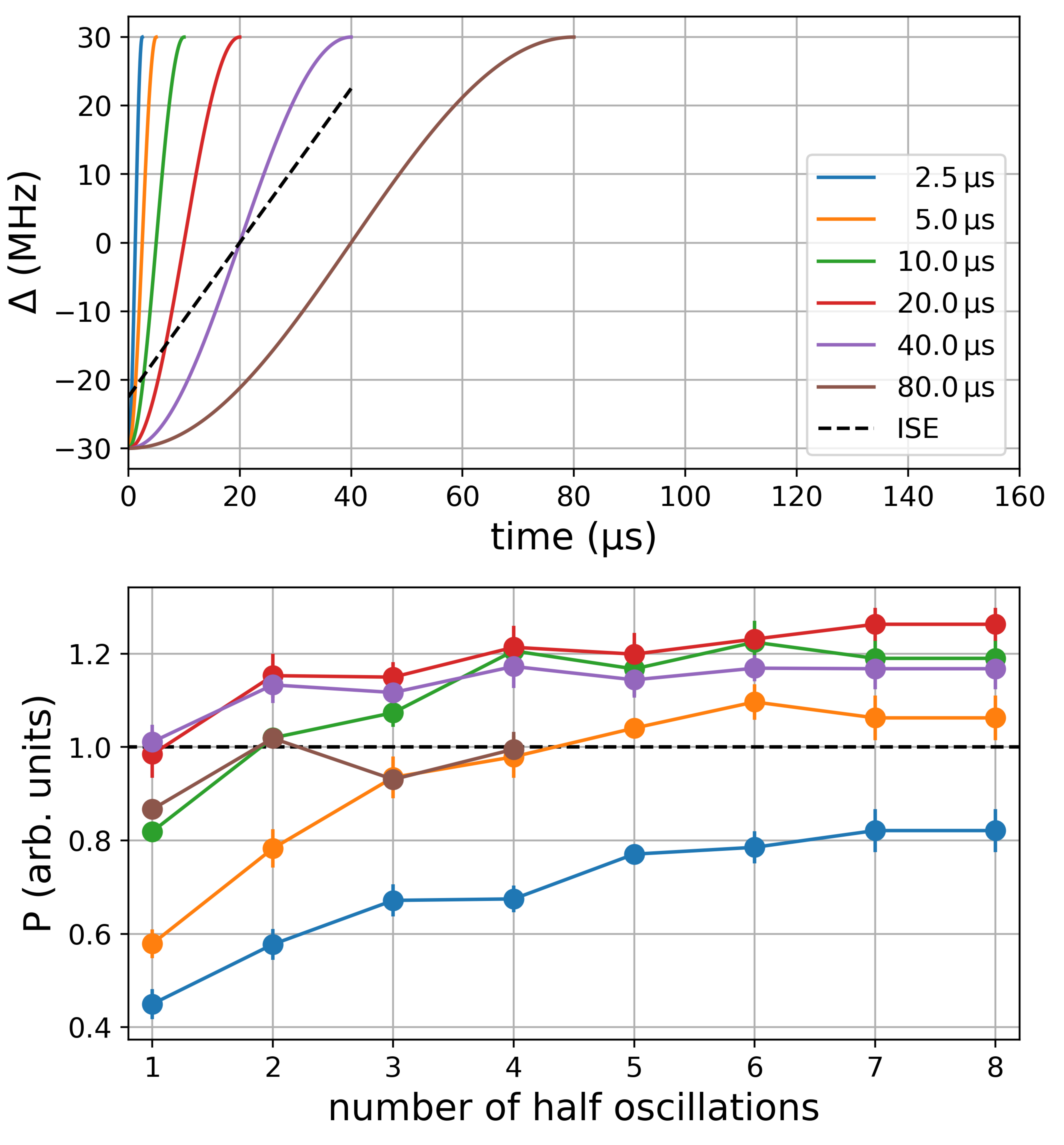}
		\includegraphics[width=0.40\textwidth]{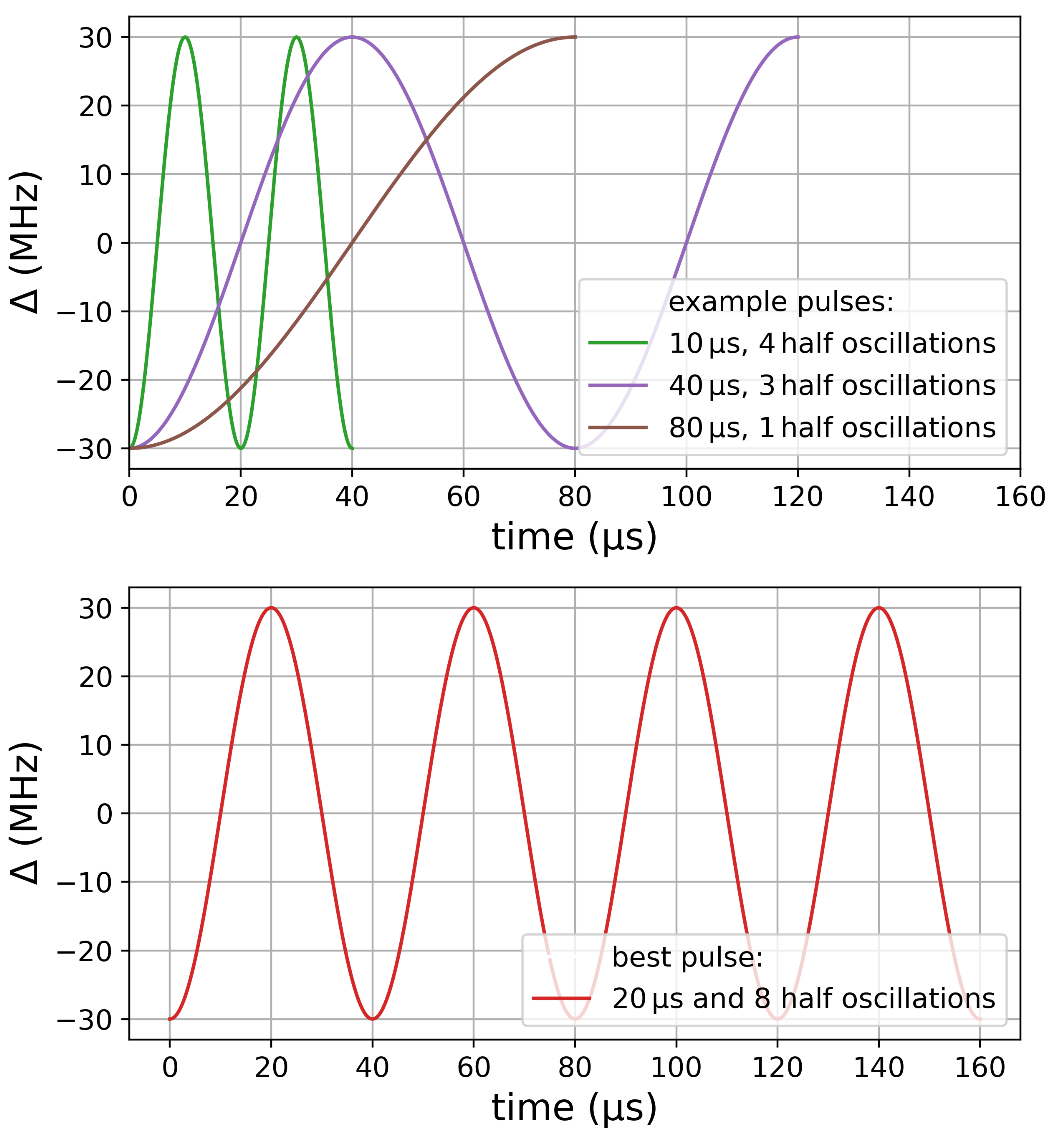}
		\caption{Comparing different sinusoidal pulses. Upper left: Half oscillations width different frequencies comparable to different speeds of the ISE-like linear sequences. Upper right: Example pulses to visualize the idea of the measurement. Lower left: For all six frequencies of the first plot, the polarization after up to eight half oscillations is measured and compared to the polarization after the linear sweep (dashed line). Lower right: Sinusoidal pulse that gave the highest polarization.}
		\label{fig:sinusoidal_comparison}
	\end{figure}
	
	Following the idea that subsequent sweeps through the resonance in alternating directions further enhance the polarization, a sinusoidal pulse is tried. To find the optimal pulse, both the frequency of the oscillation and the number of resonance passages were varied. For better comparison, the length of a half oscillation is used as a parameter instead of the frequency, which means passing through the resonance once, similar to a basic ISE-like linear sweep~(\cref{fig:sinusoidal_comparison}, upper left). The second parameter, that describes the passages through the resonance, is then given by the number of half oscillations~(examples given in \cref{fig:sinusoidal_comparison}, upper right).\\
	In~(\cref{fig:sinusoidal_comparison}, lower left) the polarization for different parameter sets (length and number of half oscillation) is compared to the standard linear sweep (dashed line). Increasing the number of resonance passages leads to an increase in polarization for all lengths. While most of the applied pulses beat the standard linear pulse, a maximum polarization for a length of 20\,$\mu$s and 8~half oscillations was found ~(\cref{fig:sinusoidal_comparison}, lower right). This pulse was characterized in the main text and used a new starting point for the sinusoidal based optimal control.
	
	
	\section{Fitted Optimized Pulse}
	\label{app:c}
	
\begin{figure}[ht]
		\centering
		\includegraphics[width = 0.5\textwidth]{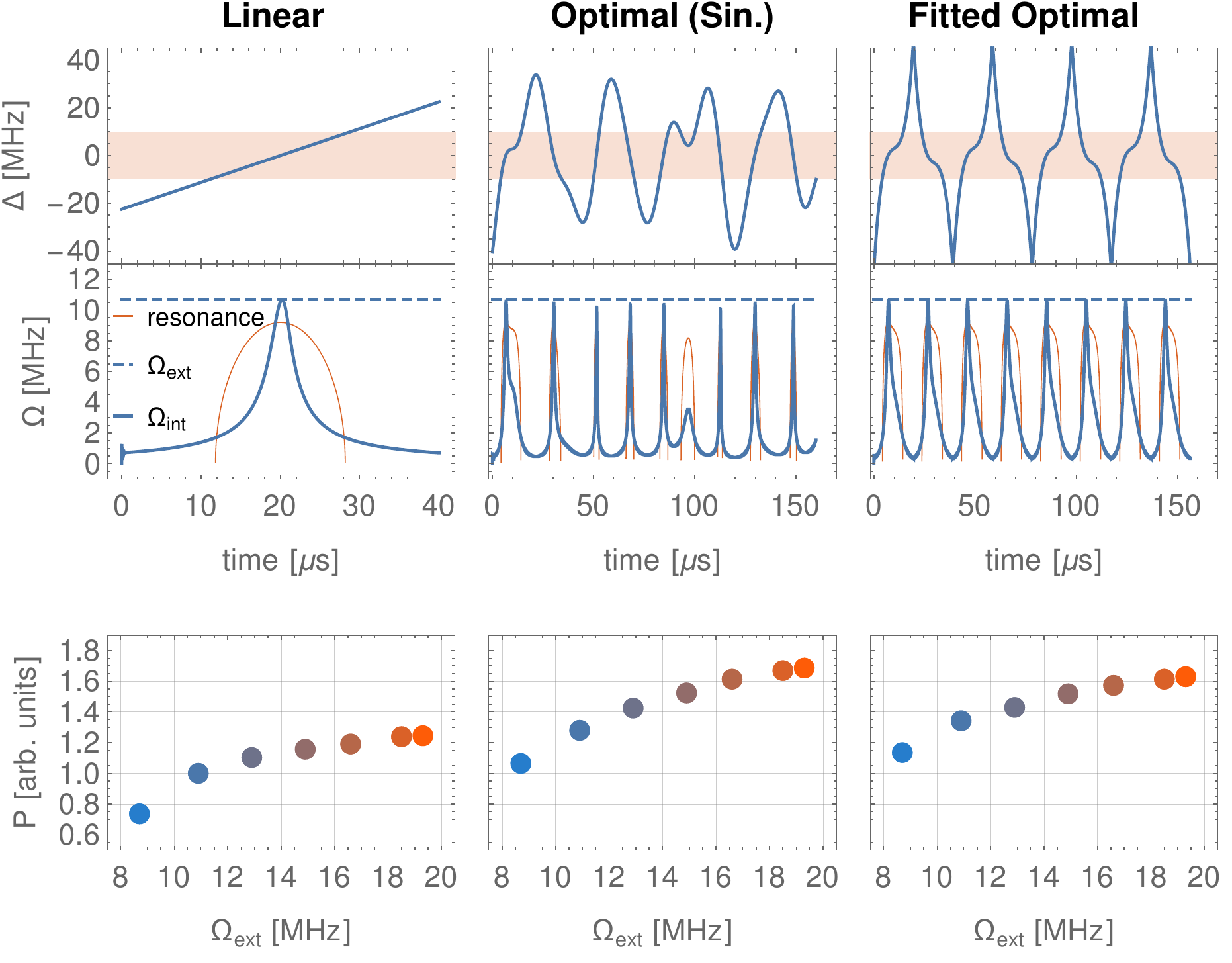}
		\caption{Polarization performance of the linear sweep (Linear), the QOC-pulse generated from a sinusoidal initial guess (Optimal (Sin.)), and its fit (Fitted Optimal) for increasing constant drive amplitudes. Guided by the outcome of the optimal control algorithm, the first $\sim20$\,$\mu$s of the optimal (sin.) pulse are re-modelled analytically by tuning polynomial functions to match the detuning's shoulder feature. Even for lower Rabi frequencies, the optimized pulses outperform the initial ISE-like linear approach.}
		\label{fig:comparison_increasing_amplitudes}
	\end{figure}

	The ``Fitted Optimal'' pulse shown on the right-hand side in \cref{fig:comparison_increasing_amplitudes} was designed by modelling the shoulder feature of the sin.-based OC pulse resulting from closed-loop optimizations of the sinusoidally varying pulse in the detuning regime. Repeating mirrored polynomial functions emulate the slow-down of the detuning sweep around resonance. A variation of the externally applied voltage kept constant during the pulses displays an improvement of polarization transfer for higher external drive amplitude. 
	Except for the highest driving amplitudes, at the limit of the experimental capabilities, the fitted pulse is equal or outperforms the pulse resulting from the closed-loop iterations. Therefore, it serves as a good starting point for the use in other, similar setups and further optimizations as well as theoretical and analytical transfer calculations and numerical simulations.
	

	\section{Na{\"i}ve Sweep Corrections}
	\label{app:d}
	While a linear sweep is applied outside the cavity, the mentioned effects lead to a non-linear sweep inside the cavity. However, the drive would appear faster when passing through the cavity resonance, compared with the part of the sweep where far outside the resonance. Measuring the cavity linewidth and Q-factor allows the calculation of the expected deviation from the linear sweep. Calculating an input function with modified amplitude that takes the cavity properties into account would be the first na{\"i}ve approach to overcome this issue. Trying this did however not improve the polarization values.
	
	\section{Coherence Measurements}
	\label{app:e}
\begin{figure}[ht]
		\centering
		\includegraphics[width=0.45\textwidth]{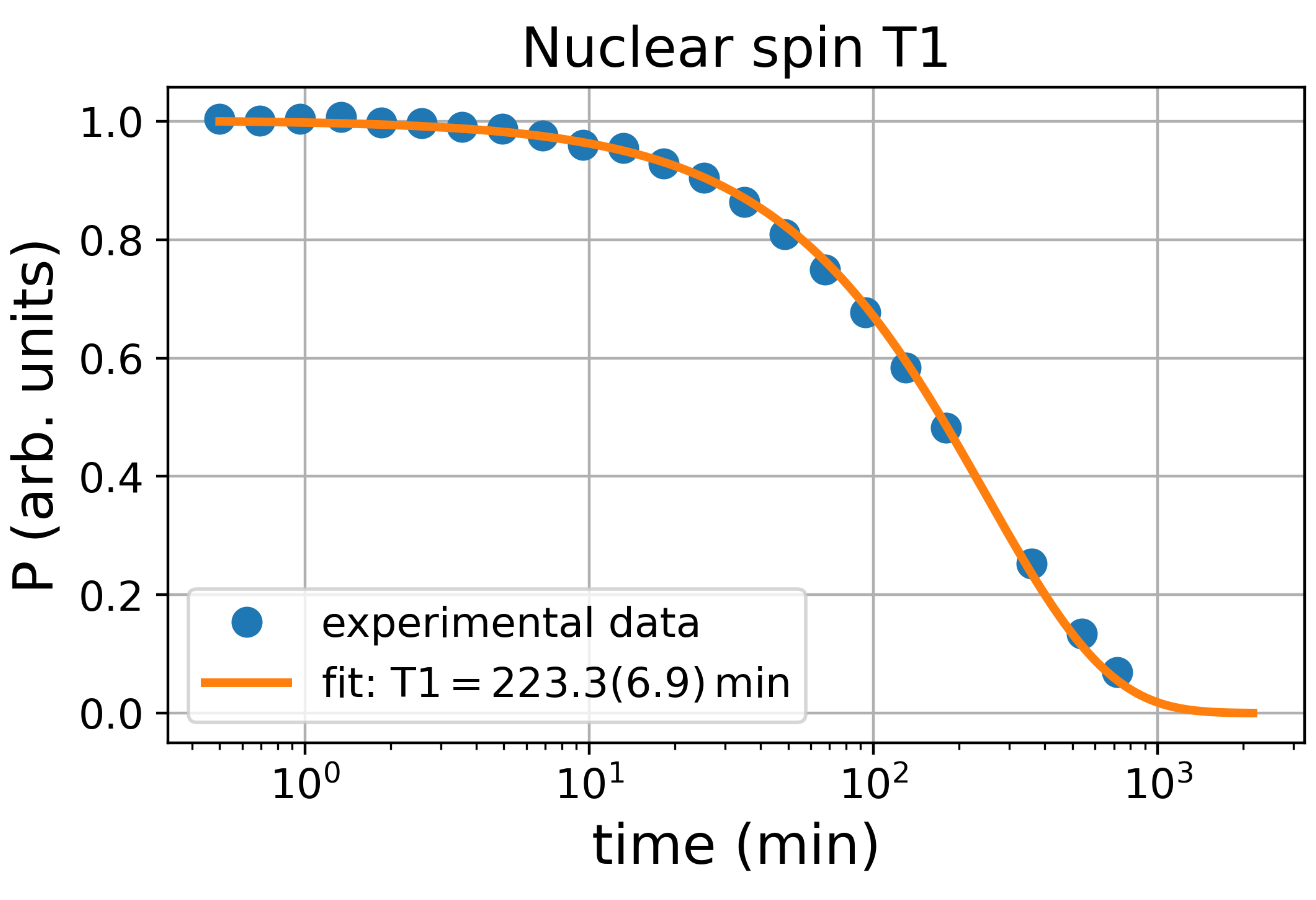}
		\includegraphics[width=0.45\textwidth]{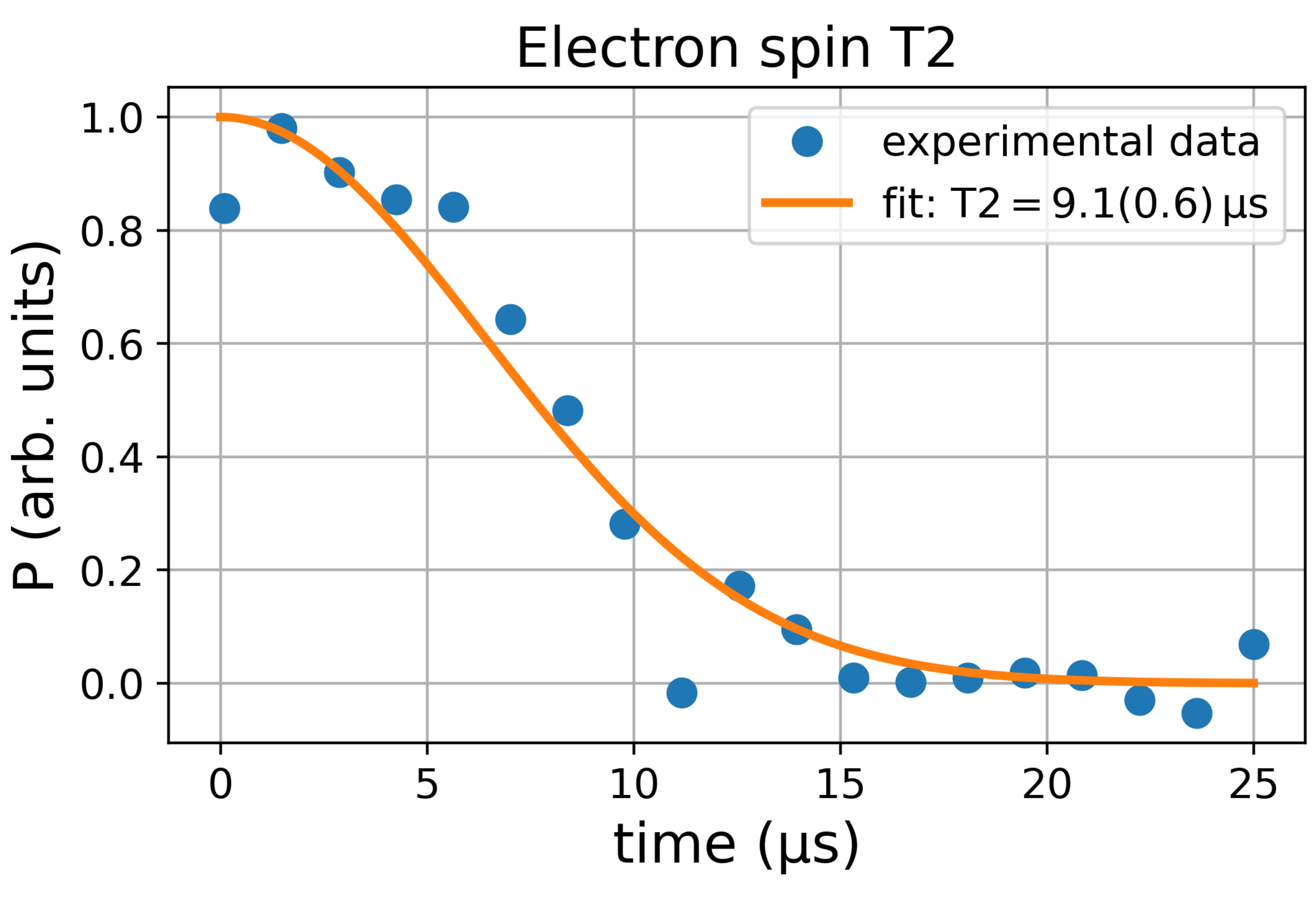}
		\caption{Left: $T_1$ measurement of the nuclear spins under experimental conditions. Right: $T_2$ measurement of the electron spin via Hahn-Echo.}
		\label{fig:coherence_measurements}
	\end{figure}

	The nuclear spin relaxation time, $T_1$, and the electron spin coherence time, $T_2$, were measured. For the nuclear spins, $T_1$ measured under experimental conditions (e.g., temperature, laser illumination) a value of around 223\,min~(\cref{fig:coherence_measurements} left) was recorded. For the $T_2$ coherence time of the electron spin, a standard Hahn-Echo sequence~\cite{hahnechopaper} was used and a value of around 9\,$\mu$s was obtained.~(\cref{fig:coherence_measurements} right).
	
	\section{Polarization Build-Up}
	\label{app:f}
	
	Polarization $p$ in the crystal is built up by iteratively applying the basic polarization sequence many times (see Fig.~3 in the main text). The final polarization will be given at the equilibrium of two competing effects: each time the basic sequence is applied, a fraction $\alpha (1-p)$ of the remaining unpolarized nuclear spins will be polarized, where $\alpha$ is the polarization power of the sequence. At the same time, the polarized nuclear spins decay at a constant rate $\gamma$.
	\begin{equation}
		\frac{dp}{dt}=\alpha (1-p) -\gamma p.
		\label{eq:pol}
	\end{equation}
	Solving \cref{eq:pol} leads to the equation
	\begin{equation}
		p(t)=p_{\mathrm{max}}\left(1- e^{-\tilde{\gamma} t}\right),
	\end{equation}
	where $p_{\mathrm{max}}=\alpha/(\gamma + \alpha)$ and the parameter $\tilde{\gamma}=\gamma+\alpha$ can be obtained from a fit to the data. From the polarization build up while applying a linear sweep $\tilde{\gamma}\approx 0.0061\,\text{min}^{-1}$ and $p_{\mathrm{max}}\approx 14140\,$arb. units. are obtained. When applying the optimal sequence $\tilde{\gamma}\approx 0.0071$ and $p_{\mathrm{max}}\approx 17850\,$arb. units. are obtained. An additional measurement T1 measurement gives $1/\gamma\approx 223\,$min~(\cref{fig:coherence_measurements} left). This translates into estimates for the final polarization of $p_{\mathrm{max}}\approx 27.8\pm 1.3\,\%$ for the linear sweep and $p_{\mathrm{max}}\approx 35.1\pm 1.7\,\%$ for the optimal sequence. Note, that the true value of $\gamma$ is probably slightly larger than in the $T_1$ measurement due to the polarization pulse sequences. If, for example, $1/\gamma\approx 200\,$min (or $1/\gamma\approx 180\,$min), is considered, the polarization for the optimal sequence drops to $p_{\mathrm{max}}\approx 26.2\pm 3.4\,\%$ (or $p_{\mathrm{max}}\approx 16.5\pm 5.2\,\%$) and similarly for the linear sweep.

	\section{Effect of the Number of Nuclei on the Model}
	\label{app:g}
	
	Each electron spin is surrounded by a large number of protons, forming the nuclear spin bath. As only a limited amount can be simulated at a time, subgroups of nuclei are considered and averaged over. Fortunately, the molecular and crystallographic properties of the naphthalene-h$_8$ specimen doped with pentacene-d$_{14}$ are well-known. Hence, the orthogonal ($A^i_{zx}$, $A^i_{zy}$) and parallel ($A^i_{zz}$) parts of the hyperfine tensor can be directly calculated~\cite{Henstra_Thesis} for the surrounding protons in the crystal. Instead of including the entire bath as a single effective nuclear spin~\cite{Wenckebach_I_2014}, several nuclei are considered individually in the polarization dynamics. This step is necessary to reflect the effects caused by complex pulse shapes obtained by the optimization, as well as the distortion by the cavity. Since the pulses cross the resonance line multiple times and are repeated successively for polarization build-up, strongly coupled nuclei are usually polarized first but can be depolarized again so that excitation is transferred to other nuclear spins. Because the electron spin is re-initialized before each pulse application, each iteration can polarize different nuclei. Up to six nuclei were considered during the initial investigation. However, for the figures presented in this manuscript, the following combinations of spins was used to keep computational resources within an acceptable range: The electron spin is coupled to three nuclear spins where the hyperfine coupling values are randomly selected from the top 30 most strongly coupled protons. This simulation is repeated and average for 1000 sets using three different random nuclei in each run. This captures the dynamics of multiple protons coupling to the electron at the same time, as well as the repetition of the transfer operation in the experiment.

	\section{Quantum Optimal Control}
	\label{app:h}
	
	Optimal control methods aim to optimize a functional $f$ by modifying time-dependent control functions $u_i(t).$ This functional is the Figure of Merit (FoM), it includes all the relevant information contributing to the quality of an operation. To simplify the optimization problem, the controls can be parameterized in terms of $N_\text{be}$ basis functions $v_\ell(t)$ with corresponding parameters $c_\ell$
	\begin{equation}
		u_i(t)=\sum_{\ell=1}^{N_\text{be}} c_\ell v_\ell(t) \, .
	\end{equation}
	
	The FoM therefore depends on the coefficients of these basis functions:
	\begin{equation}
		\text{FoM}(u_i) = f(c_\ell, v_\ell, t) \, .
	\end{equation}
	
	The solution to the problem is found using an iterative optimization algorithm, which takes in the FoM for a defined set of parameters and returns a new set of parameters. Closed-loop control involves the algorithm directly interacting with the experimental setup. Hence, it automatically takes into account real-world imperfections. Specifics of the measurement technique can be found in \cref{sec:meas_technique}.
	The algorithm improves the FoM by comparing the results from different iterations. It follows the direction of improvement, while exploring the parameter landscape and exploiting its features. An optimal set of controls is eventually obtained after a number of iterations and FoM evaluations.

	Limiting the size of the parameter landscape reduces the number of experimental runs and hence the total optimization time. The dCRAB algorithm~\cite{Doria2011, dressing_the_crab, Caneva_2011, Mueller2021decade}, in combination with the Nelder-Mead ~\cite{Gao2012,nelder1965simplex} simplex optimization algorithm, is a good choice for this.
	A small parameter space is created by randomly picking a number of basis functions, finding the optimal parameters for them, and then switching to a new basis set. An optimization in a single parameter space is called a super-iteration. This allows the optimization to start afresh and continue, even if it temporarily gets stuck in a local optimum.
	
	For the parameterization the Fourier basis is chosen, which provides a simple method to restrict the bandwidth of the controls by limiting their maximum oscillation frequency component through a capping of $\omega_{d,\ell}$ in
	\begin{equation}
		u(t) = \sum_{d=1}^{N_\text{SI}} \sum_{\ell=1}^{N_\text{be}}[A^\text{opt}_{d,\ell} \sin(\omega_{d,\ell}t)+B^\text{opt}_{d,\ell} \cos(\omega_{d,\ell} t)] \, .
	\end{equation}
	$N_\text{be}$ represents the number of basis elements (i.e., the size of the parameter space in each super-iteration), while $N_\text{SI}$ corresponds to the number of super-iterations. The parameter space of the optimization is spanned by $c_{d,\ell}=\{A^\text{opt}_{d,\ell},B^\text{opt}_{d,\ell}\}$ while $\omega_{d,\ell}$ is randomly initialized with frequencies within a pre-defined interval for each super-iteration defining the basis functions  $v_{d,\ell}=\{\sin(\omega_{d,\ell}t),\cos(\omega_{d,\ell}t)\}$. Meanwhile, the length of the pulses is kept constant.
	
	The combination of this limited search space for efficient closed-loop optimization together with the three-step ARISE protocol (see \cref{sec:protocol}) enables the encoding of a sufficient amount of information in the control pulse to substantially increase its performance~\cite{informationtheoretic}.
	
	
	\bibliography{references}
	
\end{document}